\documentclass[showpacs,superscriptaddress,amsmath,amssymb,twocolumn]{revtex4}
\usepackage{graphicx}
\usepackage{bm}

\begin{document}

\title{Energy distribution and thermodynamics of the quantum-corrected Schwarzschild black hole}

\author{Mahamat Saleh}
\email{mahsaleh2000@yahoo.fr}

\affiliation{Department of Physics, Higher Teachers' Training College, University of Maroua, P.O. Box 55, Maroua, Cameroon.}

\author{Bouetou Bouetou Thomas}
\email{tbouetou@yahoo.fr}

\affiliation{Ecole Nationale Sup$\acute{e}$rieure Polytechnique,
University of Yaounde I, P.O. Box. 8390, Cameroon}

\affiliation{The African Center of Excellence in Information and Communication
Technologies (CETIC)}

\affiliation{The Abdus Salam International Centre for Theoretical
Physics, P.O. Box 586, Strada Costiera, II-34014, Trieste, Italy}

\author{Timoleon Crepin Kofane}
\email{tckofane@yahoo.com}

\affiliation{Department of Physics, Faculty of Science, University
of Yaounde I, P.O. Box. 812, Cameroon}

\affiliation{The African Center of Excellence in Information and Communication
Technologies (CETIC)}

\affiliation{The Abdus Salam International Centre for Theoretical
Physics, P.O. Box 586, Strada Costiera, II-34014, Trieste, Italy}

\date{\today}

\begin{abstract}
In this work, energy distribution and thermodynamics are investigated in the Schwarzschild black hole spacetime when considering corrections due to quantum vacuum fluctuations. The Einstein and M{\o}ller prescriptions were used to derive the expressions of the energy in the background. The temperature and heat capacity were also derived. The results show that due to the quantum fluctuations in the background of the Schwarzschild black hole, all the energies increase and Einstein energy  differs from M{\o}ller's one. Moreover, when increasing the quantum correction factor ($a$), the difference between Einstein   and M{\o}ller energies, the Unruh-Verlinde temperature as well as the heat capacity of the black hole increase while the Hawking temperature remains unchanged.
\end{abstract}


\maketitle

The localization of energy and momentum in general relativity is one of the oldest problems in gravitation which still lacks a definite answer. The key point for the issue is whether the
energy of spacetime may be localized or not. It was Sarracino and Cooperstock and other researchers that pointed the possibility that the energy is localizable\cite{cs}. Through the years, many efforts were made to develop truly invariant general prescription for gravitational field energy localization\cite{R1,R2,R3,R4,R5,R6,R7}.
Using the various prescriptions, many researchers derived energy and momentum expressions in various spacetimes\cite{cs,xulu2,rad, vag,vag2,lali,vag3,virb1, virb2, virb3, aguir,xul, rad1,rad2, rad3,xulubt,msbk}.  The obtained results show that for general Kerr-Schild class of spacetimes, the expressions of the energy coincide for all of the prescriptions except for the M{\o}ller's prescription. This divergence guides such investigation leading to the choice of an appropriate result for such spacetime\cite{wei}.
Theoretical studies on black holes have attracted substantial attention since the advent of general relativity. The investigation of energy distribution in black holes spacetimes is one of the hot topics in modern physics.

Black holes are thermal systems, radiating as black bodies with characteristic temperatures
and entropies. During the past 40 years, research in the theory of black holes in general
relativity has brought to light strong hints of a very deep and fundamental
relationship between gravitation, thermodynamics, and quantum theory.
The cornerstone of this relationship is black hole thermodynamics, where it
appears that certain laws of black hole mechanics are, in fact, simply the
ordinary laws of thermodynamics applied to a system containing a black hole\cite{wald}.
All over the last decade, there have been several outstanding approaches toward a
statistical mechanical computation of the Bekenstein-Hawking (BH) entropy\cite{carlip1, carlip2, stromva, prl80}.
Black holes as thermodynamical systems are widely found in the literature\cite{haw, haw2, bek2, pete, fabnag, zhao2hu, gibhaw, carlip1, carlip2, stromva, prl80, pw2, bacaha, mbk}.

We know that the vacuum undergoes quantum fluctuations. This phenomenon is the appearance of energetic particles out of nothing, as allowed by the uncertainty principle. Vacuum fluctuations have observable consequences like Casimir force between two plates in vacuum. Due to quantum fluctuations, the evolution of slices in the black hole geometry will lead to the creation of particle pairs which facilitates black holes' radiation\cite{mathur}.

Recently, Wontae and Yongwan\cite{wontae} investigated phase transition of the quantum-corrected Schwarzschild black hole and concluded that there appear a type of Grass-Perry-Yaffe phase transition due to the quantum vacuum fluctuations and this held even for the very small size black hole. More recently, we investigate quasinormal modes of a quantum-corrected Schwarzschild black hole and show that the scalar field damps more slowly and oscillates more slowly due to the quantum fluctuations\cite{ms, ms2}.

In this paper, the energy distribution and thermodynamics of a
quantum-corrected Schwarzschild black hole are investigated in order to highlight the energetic and thermodynamical behaviors of the black hole when the vacuum fluctuations are taken into account.

According to the work of Kazakov and Solodukhin on quantum deformation of the Schwarzschild
solution \cite{kazsol}, the background metric of the Schwarzschild black hole is defined by
\begin{equation}\label{01}
    ds^2=f(r)dt^{2}-f(r)^{-1}dr^{2}-r^{2}(d\theta^{2}+sin^2\theta d\varphi^2),
\end{equation}
where
\begin{equation}\label{2}
    f(r)=-\frac{2M}{r}+\frac{1}{r}\int^r U(\rho)d\rho.
\end{equation}

For an empty space, $U(\rho)=1$. Thus we obtain the Schwarzschild
metric
\begin{equation}\label{3}
ds^{2}=\big(1-\frac{2M}{r}\big)dt^{2}-\big(1-\frac{2M}{r}\big)^{-1}dr^{2}-r^{2}(d\theta^{2}+sin^{2}\theta
d\varphi^{2}),
\end{equation}
where $M$ is the black hole mass.
The event horizon of the black hole is localized at $r_{EH}=2M$.

Taking into account the quantum fluctuation of the vacuum, the
quantity $U(\rho)$ transforms to \cite{wontae}
\begin{equation}\label{4}
    U(\rho)=\frac{e^{-\rho}}{\sqrt{e^{-2\rho}-\frac{4}{\pi}G_R}},
\end{equation}
where $G_R=G_Nln(\mu/\mu_0)$, $G_N$ is the Newton constant and $\mu$ is a scale parameter.

The background metric of the quantum-corrected Schwarzschild black hole can then be read as

\begin{equation}\label{5}
ds^{2}=f(r)dt^{2}-f(r)^{-1}dr^{2}-r^{2}(d\theta^{2}+sin^{2}\theta
d\varphi^{2}),
\end{equation}
 with $f(r)=\big(-\frac{2M}{r}+\frac{\sqrt{r^{2}-a^{2}}}{r}\big)$ and $a^2=4G_R/\pi$.
 The event horizon of such black hole is located at the radius $r_{EH}=\sqrt{4M^2+a^2}$. It is clear that the area of the event horizon increases with the quantum-correction parameter, $a$.

 The energy-momentum complex from M{\o}ller's prescription can be
derived as\cite{vag, vag3}:
\begin{equation}\label{e5}
    \tau^{\mu}_{\nu}=\frac{1}{8\pi}\varepsilon^{\mu\lambda}_{\nu,
    \lambda},
\end{equation}
where the superpotentials are
\begin{equation}\label{e6}
    \varepsilon^{\mu\lambda}_{\nu}=\sqrt{-g}(\frac{\partial g_{\nu\sigma}}{\partial
    x^{k}}-\frac{\partial g_{\nu k}}{\partial x^{\sigma}})g^{\mu
    k}g^{\lambda \sigma}.
\end{equation}

The energy of the physical system in a four-dimensional background
is given by
\begin{equation}\label{e7}
    E_{M}=\int\int\int\tau^{0}_{0}dx^{1}dx^{2}dx^{3}.
\end{equation}

Substituting equation(~\ref{5}) in (~\ref{e6}), the M{\o}ller's
energy distribution for the quantum-corrected Schwarzschild black hole can be written as:
\begin{equation}\label{e8}
    E_{M}=M+\frac{a^{2}}{2\sqrt{r^2-a^2}}.
\end{equation}

The energy distribution from Einsten's energy-momentum complex
is\cite{R1,vag3}:
\begin{equation}\label{e9}
    \theta^{\mu}_{\nu}=\frac{1}{16\pi}h^{\mu l}_{\nu, l},
\end{equation}
where \begin{equation}\label{e10}
    h^{\mu l}_{\nu}=-h^{l \mu}_{\nu}= \frac{g_{\nu
    n}}{\sqrt{-g}}[-g(g^{\mu n}g^{l m}-g^{l n}g^{\mu m})]_{,m}.
\end{equation}

Since Einstein's energy-momentum complex is restricted to
quasi-cartesian coordinates, we can express the above
metric(~\ref{5}) in quasi-cartesian coordinates defined as:
\begin{eqnarray}\label{e11}
   \left\{
    \begin{array}{lcl}
      T &=& t+r - \int^{ }_{ }{f(r)^{-1}dr} \\
      x &=& r sin\theta cos\varphi \\
      y &=& r sin\theta sin\varphi \\
      z &=& r cos\theta
    \end{array}\right.
\end{eqnarray}

The metric can then be expressed as:
\begin{equation}\label{e12}
    \begin{array}{lcl}
    ds^{2}&=&-dT^{2}+dx^{2}+dy^{2}+dz^{2}\\
    &&+(1-f(r))\bigg[dT-\frac{xdx+ydy+zdz}{r}\bigg]^{2}.
    \end{array}
\end{equation}

The energy of the physical system is, once again, given by the
formula
\begin{equation}\label{e13}
    E_{E}=\int\int\int\theta^{0}_{0}dx^{1}dx^{2}dx^{3}.
\end{equation}

$\theta^{0}_{0}$ is evaluated using equations (~\ref{e9}),
(~\ref{e10}) and (~\ref{e12}) and substituted in (~\ref{e13}) to get
the energy

\begin{equation}\label{14}
    E_{E}=M-\frac{1}{2}\sqrt{r^2-a^2}+\frac{r}{2}.
\end{equation}

\begin{figure}[h!]
\begin{center}
  \includegraphics[width=7.5cm]{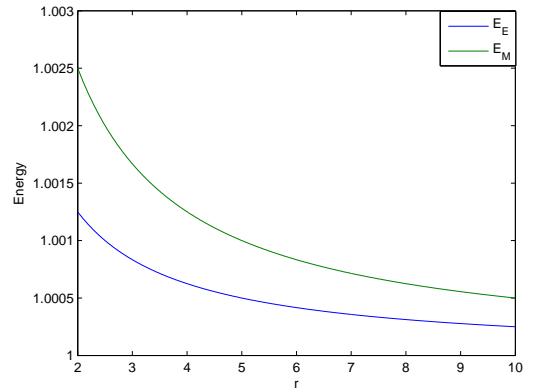}\\
  \caption{Behavior of Einstein and M{\o}ller energies versus the radial position $r$ for $M=1$.}\label{fige1}
\end{center}
\end{figure}

\begin{figure}[h!]
\begin{center}
  \includegraphics[width=7.5cm]{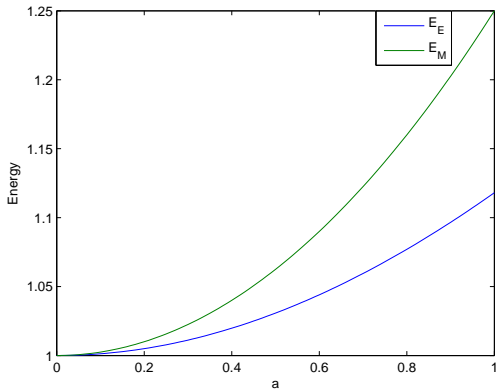}\\
  \caption{Behavior of Einstein and M{\o}ller energies versus the quantum correction parameter $a$ for $M=1$ and $r=r_{EH}$.}\label{fige2}
\end{center}
\end{figure}

The behavior of these energies is represented on Figs. \ref{fige1} and \ref{fige2}.

These figures show that the energies decrease when increasing $r$ and for fixed $r$, Einstein energy is greater than M{\o}ller energy. At the event horizon, $r=r_{EH}=\sqrt{4M^2+a^2}$, the two energies coincide when $a=0$ and, for $a\ne 0$, the gap between the two energies increases when increasing $a$.

The temperature of the black hole is given by the Unruh-Verlinde matching\cite{Verlinde, Unruh, Konoplya, Liuwangwei}
\begin{equation}\label{t1}
    T=\frac{\hbar}{4\pi r^2}(2M+\frac{a^2}{\sqrt{r^2-a^2}})
\end{equation}

\begin{figure}[h!]
\begin{center}
  \includegraphics[width=7.5cm]{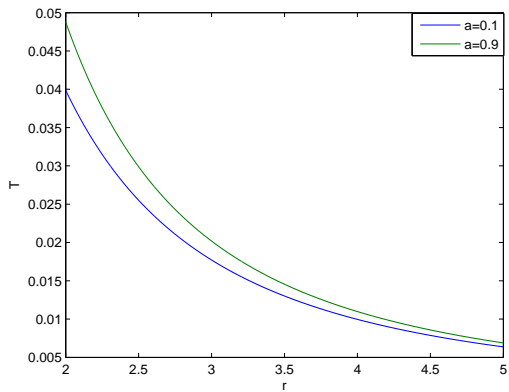}\\
  \caption{Variation of the temperature versus the radial position $r$ for different values of the quantum-correction parameter $a$.}\label{figt1}
\end{center}
\end{figure}

Its behavior is represented on Fig. \ref{figt1}.

Through this figure, we can see that the temperature decreases when increasing $r$. Moreover, when increasing the quantum-correction parameter $a$, the temperature increases.

The Hawking temperature is the temperature at the event horizon. The horizon radius $r_H$ of the quantum-corrected Schwarzschild black hole is given by
\begin{equation}\label{t01}
    r_H=\sqrt{4M^2+a^2}.
\end{equation}

Using this expression, the Hawking temperature reads
\begin{equation}\label{t2}
    T_H=T|_{r=r_H}=\frac{\hbar}{8\pi M},
\end{equation}
which corresponds to the Hawking temperature of the Schwarzschild black hole free from any kind of correction.

The entropy of the black hole is given by
\begin{equation}\label{t02}
    S=\frac{A}{4}=\pi r_H^2=4\pi M^2+\pi a^2.
\end{equation}

Substituting Eq. (\ref{t02}) into Eq. (\ref{t2}) yields
\begin{equation}\label{t3}
    T_H=\frac{\hbar}{4\pi}\left(\frac{S}{\pi}-a^2\right)^{-\frac{1}{2}}.
\end{equation}

The heat capacity of the black hole is
\begin{equation}\label{t4}
    C=T_H\left(\frac{\partial S}{\partial T_H}\right)=-2S+2\pi a^2.
\end{equation}
We can clearly see that the quantum vacuum fluctuations contribute to the heat capacity of the black hole by a positive quantity $C_a=2\pi a^2$.

In summary, we have studied energy distribution and thermodynamics for the Schwarzschild black hole when considering quantum corrections due to the quantum vacuum fluctuations. Concerning energy distribution, we derived the energy from Einstein and M{\o}ller prescriptions. Their behaviors show that when $a=0$, the two energies coincide $(E_E=E_M=M)$. When introducing the quantum-correction to the black hole, the M{\o}ller energy becomes greater than the Einstein energy, and the difference increases when increasing the quantum-correction parameter $a$ (see Figs. \ref{fige1} and \ref{fige2}). The event horizon radius of the black hole as well as its Unruh-Verlinde temperature increase with the quantum-correction parameter. The Hawking temperature stays unchanged although the horizon changes. The heat capacity increases with the quantum-correction parameter. Thus we can conclude that quantum vacuum fluctuations generate positive energy increasing the energy of the black hole and help stabilizing thermodynamically the black hole. At the event horizon, quantum vacuum fluctuations increase the energies (see Fig. \ref{fige2}) without changing the Hawking temperature which remains proportional to the inverse of the black hole mass.


\begin{thebibliography}{99}
\bibitem{cs} Cooperstock F I  and Sarracino R S 1978 \emph{J. Phys. A: Math. Gen.} \textbf{11} 877.
\bibitem{R1} Einstein A 1915 \emph{Preuss. Akad. Wiss. Berlin} \textbf{47} 778.
\bibitem{R2} Papapetrou A 1948 \emph{Proceedings of the Royal Irish Academy A} \textbf{52} 11.
\bibitem{R3} Tolman R C 1930 \emph{Phys. Rev.} \textbf{35} 875
\bibitem{R4} Landau L D and Lifschitz E M 1951 \emph{The Classical Theory of Fields} (Addison-Wesley Press, Reading, Massachusetts., New York) p.317.
\bibitem{R5} Bergmann P G and Thomson R 1953 \emph{Phys. Rev.} \textbf{89} 400.
\bibitem{R6} Weinberg S 1972 \emph{Gravitation and Cosmology: Principles and Applications of General Theory of Relativity} (John Wiley and Sons, Inc.,
New York) p.165.
\bibitem{R7} Goldberg J N 1958 \emph{Phys. Rev.} \textbf{111} 315.
\bibitem{xulu2} Xulu S S 2007  \emph{Int. J. Theor. Phys.} \textbf{46} 2915.
\bibitem{rad} Radinschi I and Grammenos T 2008  \emph{Int. J. Theor. Phys.} \textbf{47} 1363.
\bibitem{vag} Vagenas E C 2006 \emph{Mod. Phys. Lett.} A \textbf{21} 1947.
\bibitem{vag2} Vagenas E C 2004 \emph{Mod. Phys. Lett.} A \textbf{19} 213.
\bibitem{lali} Xulu S S,  arXiv:gr-qc/0304081v1.
\bibitem{vag3} Vagenas E C 2003 \emph{Int. J. Mod. Phys.} A \textbf{18} 5949.
\bibitem{virb1} Virbhadra K S 1990 \emph{Phys. Rev.} D \textbf{41} 1086.
\bibitem{virb2} Virbhadra K S 1990 \emph{Phys. Rev.} D \textbf{42} 2919.
\bibitem{virb3} Virbhadra K S 1990 \emph{Phys. Rev.} D \textbf{42} 1066.
\bibitem{aguir} Aguirregabiria J M, Chamorro A and Virbhadra K S 1996 \emph{Gen. Rel. Grav.} \textbf{28} 1393.
\bibitem{xul} Xulu S S 1998 \emph{Int. J. Mod. Phys.} D \textbf{7} 773.
\bibitem{rad1} Radinschi I 1999 \emph{Acta Physica Slovaca} \textbf{49} 789.
\bibitem{rad2} Radinschi I 2000 \emph{FIZIKA} B \textbf{9} 43.
\bibitem{rad3} Radinschi I 2000 \emph{Mod. Phys. Lett.} \textbf{15} 803.
\bibitem{xulubt} Xulu S S 2007 \emph{Int. J. Theor. Phys.} \textbf{46} 2915.
\bibitem{msbk} Mahamat S, Bouetou B T and Kofane T C 2011 \emph{Commun. Theor. Phys.} \textbf{55} 291.
\bibitem{wei} Wei Y-H 2008 \emph{Chin. Phys. Lett.} \textbf{25} 2782.
\bibitem{wald} Wald R M 2001 \emph{Living Rev. Rel.} \textbf{4} 6.
\bibitem{carlip1} Carlip S 1995 \emph{Phys. Rev.} D \textbf{51} 632.
\bibitem{carlip2} Carlip S 1997 \emph{Phys. Rev.} D \textbf{55} 878.
\bibitem{stromva} Strominger and Vafa 1996 \emph{Phys. Lett.} B \textbf{379} 99.
\bibitem{prl80} Ashtekar A, Baez J, Corichi A and Krasnov K 1998 \emph{Phys. Rev. Lett.} \textbf{80} 904.
\bibitem{haw} Hawking S W 1975 \emph{Commun. Math. Phys.} \textbf{43} 199.
\bibitem{haw2} Hawking S W 1974 \emph{Nature} \textbf{30} 248.
\bibitem{bek2} Bekenstein J D 1973 \emph{Phys. Rev.} D \textbf{7} 2333.
\bibitem{pete} Pete M, arXiv:physics.pop-ph/0906.4849v1.
\bibitem{fabnag} Farmanya A, Abbasi S and Naghipour A 2008 \emph{Acta Phys. Pol.} A \textbf{114} 651.
\bibitem{zhao2hu} Zhao R, Zhao H-X and Hu S-Q 2007 \emph{Mod. Phys. Lett.} A \textbf{22} 1737.
\bibitem{gibhaw} Gibbons G W and Hawking S W 1977 \emph{Phys. Rev.} D \textbf{15} 2738.
\bibitem{pw2} Parikh M K and Wilczek F 1998 \emph{Phys. Rev.} D \textbf{58} 064011.
\bibitem{bacaha} Bardeen J M, Carter B and Hawking S W 1973 \emph{Comm. Math. Phys.}  \textbf{31} 161.
\bibitem{mbk} Mahamat S,  Bouetou B T and Kofane T C 2012 \emph{Gen. Relativ. Gravit.} \textbf{44} 2181.
\bibitem{mathur} Mathur S D 2009 \emph{Class. Quantum Grav.} \textbf{26} 224001.
\bibitem{wontae} Wontae K and Yongwan K 2012 \emph{Phys. Lett.} B \textbf{718} 687.
\bibitem{ms} Mahamat S, Bouetou B T and Kofane T C 2014 \emph{Astrophys. Space Sci.} \textbf{350} 721.
\bibitem{ms2} Mahamat S., Bouetou B.T. and Kofane T C 2016 \emph{Astrophys. Space Sci.}  \textbf{361} 137.
\bibitem{kazsol} Kazakov  D I and Solodukhin S N 1994 \emph{Nucl. Phys.} B \textbf{429} 153.
\bibitem{Verlinde} Verlinde E 2011 \emph{JHEP} \textbf{04} 029.
\bibitem{Unruh} Unruh W G 1976 \emph{Phys. Rev.} D \textbf{14} 870.
\bibitem{Konoplya} Konoplya R A 2010 \emph{Eur. Phys. J.} C  \textbf{69} 555.
\bibitem{Liuwangwei} Liu Y-X, Wang Y-Q and Wei S-W 2010 \emph{Class. Quantum Grav.} \textbf{27} 185002.

\end{thebibliography}
\end{document}